\documentclass[conference]{IEEEtran}

\usepackage{cite}
\usepackage{amsmath,amssymb,amsfonts}
\usepackage{algorithmic}
\usepackage{graphicx}
\usepackage{textcomp}
\usepackage{xcolor}
\def\BibTeX{{\rm B\kern-.05em{\sc i\kern-.025em b}\kern-.08em
    T\kern-.1667em\lower.7ex\hbox{E}\kern-.125emX}}

\begin{document}
\makeatletter
    \newcommand{\linebreakand}{%
      \end{@IEEEauthorhalign}
      \hfill\mbox{}\par
      \mbox{}\hfill\begin{@IEEEauthorhalign}
    }
\makeatother

\title{Secure Decentralized Online Gaming\\with Lending Functionalities}

\author{\IEEEauthorblockN{Katharina Alefs}
\IEEEauthorblockA{\textit{Department of Computer Science} \\
\textit{INSA Lyon}\\
69621 Villeurbanne, France\\
katharina.alefs@insa-lyon.fr}
\and
\IEEEauthorblockN{Florian Hartl}
\IEEEauthorblockA{\textit{Department of Computer Science} \\
\textit{INSA Lyon}\\
69621 Villeurbanne, France\\
florian.hartl@insa-lyon.fr}
\and
\IEEEauthorblockN{Luke Newman}
\IEEEauthorblockA{\textit{Department of Computer Science} \\
\textit{INSA Lyon}\\
69621 Villeurbanne, France\\
luke.newman@insa-lyon.fr}
\linebreakand 
\IEEEauthorblockN{Banu Özdeveci}
\IEEEauthorblockA{\textit{Department of Computer Science} \\
\textit{INSA Lyon}\\
69621 Villeurbanne, France\\
banu.ozdeveci@insa-lyon.fr}
\and
\IEEEauthorblockN{Wisnu Uriawan}
\IEEEauthorblockA{\textit{LIRIS Laboratory UMR 5205 CNRS}\\
\textit{INSA Lyon}\\
69621 Villeurbanne, France\\
wisnu.uriawan@insa-lyon.fr}
}

\maketitle

\begin{abstract}
We present a decentralized online gaming platform 
implemented as a Decentralized Application (DApp) on the Ethereum blockchain. The gaming platform enables secure gaming, where the account balances and the stakes of the players are secured by a smart contract. Moreover, the fair enforcement of the game rules and the deposit of the winnings of the players and the gaming platform into their accounts are guaranteed by the smart contract. The gaming platform proposes lending functionalities that allow players to securely borrow tokens from the gaming platform in order to participate in the games.
\end{abstract}


\section{Introduction}
\label{sec:introduction}

Online gaming comprising of probability games is a popular application. Examples of probability games include dice roll, slots, roulette, and blackjack. These games are also called games of chance. In this paper, we present a gaming platform implemented as a Decentralized Application (DApp) on Ethereum. The gaming platform also provides lending functionalities. After the initialization of the gaming platform by the owner, users are able to interact with the main smart contract and the various functionalities embedded within it. We propose to implement a token system, which means that users must buy tokens to partake in the functionalities within the game. The smart contract offers a wide range of use cases, such as buying tokens (in the game currency), playing different games, withdrawing tokens to convert back to Ether, and lending tokens to the users.

The advantage of implementing the application as a DApp with smart contracts is that it is decentralized. This decentralization implies that no external third parties can control the application or intervene with the smart contract. As opposed to existing centralized online gaming websites, a third party is not able to declare new unfair rules, fees, or any other parameters to take advantage of users. The only interactions with our gaming smart contract are between individual users and the gaming platform owner.

The gaming scenario is very suitable for applying smart contracts since both parties can trust in secure and reliable transactions and a fair game is established. Upon interaction between a user account and our smart contract, there is a guarantee that the user can withdraw whenever they wish after either winning or losing in the game, which also respects the fundamentals of a smart contract and decentralization. In addition, users who interact with the game smart contract, remain anonymous since the only given identifier of a user is their account ID.

Our application is built using Remix, which we use to write, compile, debug and deploy solidity files. For front-end interaction we use React (javascript) to provide a visual UX for the smart contract. In order to compile our solidity files into code that could be run on our client-side application (React), we used Hardhat. To enable interactions with our smart contract on the blockchain, we use Ether.js. On top of all this, we use Metamask to help with managing accounts on the Ethereum blockchain.

The remainder of the paper is structured as follows: Section \ref{sec:related-work} discusses related work. Section \ref{sec:probability-games} describes probability games. Section \ref{sec:use-cases} presents the use cases of our gaming platform. Section \ref{sec:gaming-dapp-design} presents the design of our gaming platform DApp. Section \ref{sec:implementation} discusses implementation details and lists the source code of some of the main functions of the smart contract. Section \ref{sec:conclusion} concludes the paper.

\section{Related Work}
\label{sec:related-work}

In this section, we look at some of the existing blockchain-based online gaming platforms. To the best of our knowledge, none of these platforms provide lending functionalities at the moment for fungible tokens. In contrast, our gaming platform proposes the possibility of loans to the players.

FUNToken.io \cite{funtoken} is an Ethereum-based gaming platform that offers a large user community. They plan to supplement on-chain transactions with side-chain transactions for lower cost and lower latency. The Atari Token \cite{atari} has good integration capabilities with other decentralized applications. Atari Token provides services to individuals as well as businesses, supports multiple platforms, offers liquidity guarantees, and easy payments.

CoinPoker \cite{poker} is a blockchain technology-based platform that uses USDT stable coin as the main in-game currency and the CHP cryptocurrency for bonuses. It offers instant and secure transactions using USDT, ETH, BTC, and CHP tokens. The platform allows anonymity with no KYC checks. The BetU platforms \cite{betu} include BetU Verse, EarnU, and BetU. It provides a virtual reality platform, prediction games (sports and esports), and betting platform. BETU tokens are utilized for all winnings, betting rewards, incentives, staking, burning, whale holder benefits, purchasing of NFTs, and governance of the platforms.

LOTTO (The Immutable World Lottery) is a decentralized and cross-border lottery game. The LOTTO \cite{lotto} lottery runs continuously without a third party. The protocol produces a provably-fair winner using an oracle service. Dotmoovs \cite{moov} is a peer-to-peer game playing platform based on an artificial intelligence system. The Moov token is an asset that supports all transactions including buying, selling, and renting.

Exeedme \cite{exeedme} allows gamers to engage and interact with their favorite games. Exeedme provides an earning environment to several gaming communities with one token to fuel the entire platform. Wagerr \cite{wagerr} is a permissionless blockchain-based platform that offers sports betting. Wagerr is supported by Oracles and provides transparency. Wagerr cryptocurrency is designed with a deflationary mechanism that helps the coin retain value under various conditions to protect from market drops and guard price to equilibrium over time.

\section{Probability Games} \label{sec:probability-games}
The outcome of a probability game \cite{Alexis,Harvard} depends on the likeliness of certain events to occur. Probability games may also be called games of chance. In this section, we describe some of the probability games that could be played on our gaming platform.

\subsection{Roll the Dice}
In this game, the player enters a guess and a stake. She loses or wins 6 times the stake.

\subsection{Slot machine}
The player enters the stake. The result is three digits. If two digits are the same, \textit{Reward = 3 $\times$ stake}. If three digits are the same, \textit{Reward = 9 $\times$ stake}.

\subsection{Roulette}
The player enters the stake and guesses a number and a color. If the color is correct, \textit{Reward = 2 $\times$ stake}. If the number is also correct, \textit{Reward = 36 $\times$ stake}.

\subsection{Blackjack}
\begin{itemize}
\item \textit{Goal:} Get more points than the dealer, but at most 21 points.
\item Draw 2 cards for yourself and 1 for the dealer.
\item \textit{Simpliﬁcation:} Ace is 11 points and unlimited cards.
\item Draw as many cards as you want
(one after another).
\item If you have more than 21 points, you lose directly.
\item The dealer draws cards until they have at least 16 points.
\item If the dealer has more than 21 points, you win directly.
\item If you have more points than the dealer, you win.
\item \textit{Reward:} Twice your stake
\end{itemize}

\section{Use Cases} \label{sec:use-cases}

The use case diagram in Fig. \ref{fig:usecases} demonstrates the different use cases of the application and who out of the gaming platform owner and user can interact with each use case. The order in which each use case should follow the other is the order in which the use cases are presented in the diagram (top to bottom).

\begin{figure}[!htb]
\includegraphics[width=\columnwidth]{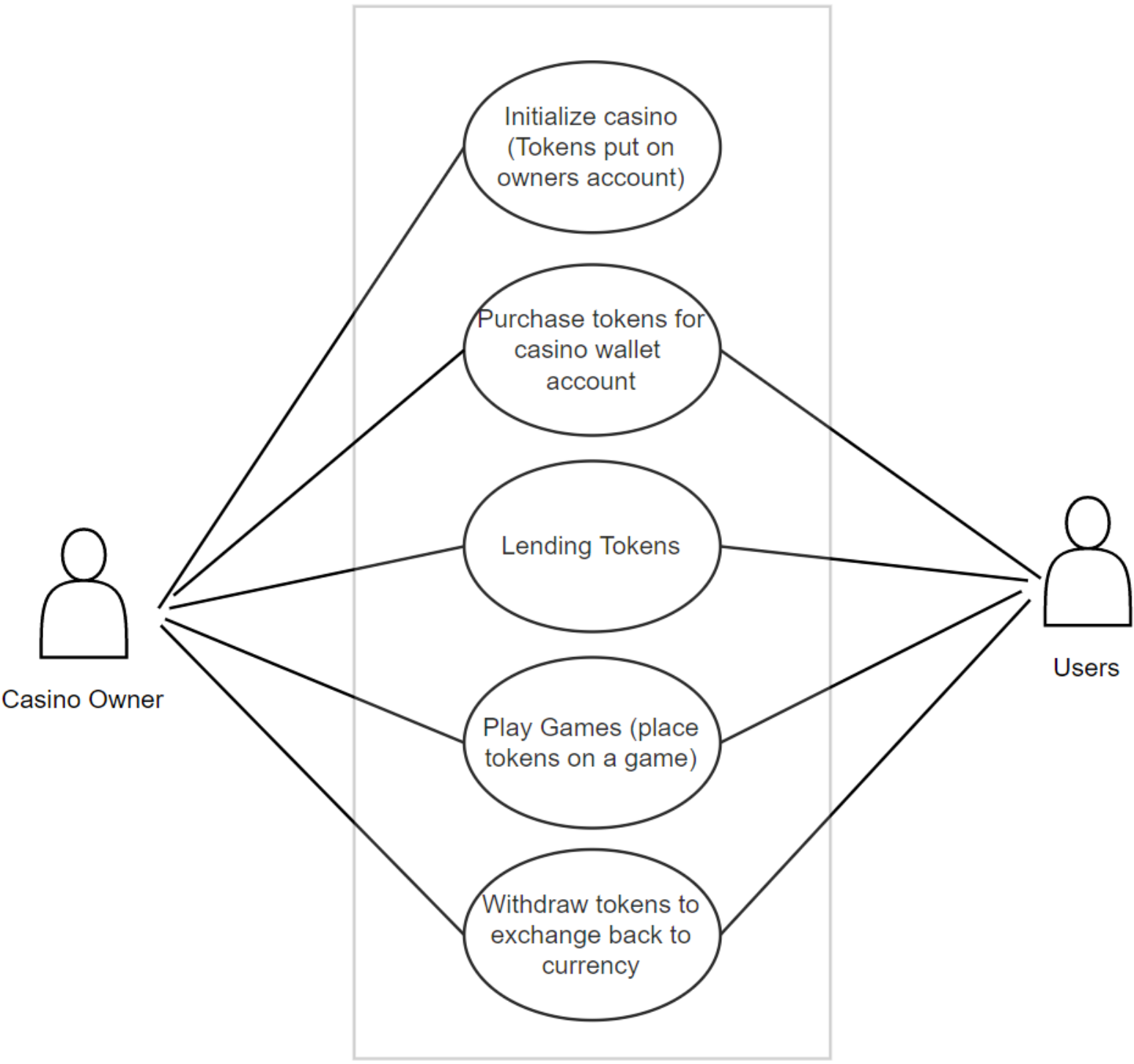}
\caption{Use Cases}
\label{fig:usecases}
\end{figure}

\begin{enumerate}
\item \textbf{Initialize the gaming platform:} The owner must first initialize the gaming platform by investing some amount of Ether for there to be a ``gaming platform bank''. The users can win tokens from the bank or pay tokens to the bank based on the result of their game. Once this is done, the gaming platform is ready to be interacted with by the users.

\item \textbf{Purchase tokens:} For this use case, a user can buy tokens using their existing account balance from the Ethereum network. Once a user purchases tokens, an account balance and user ID is generated within the gaming platform and the user is ready to play the games. The gaming platform owner can also top-up on tokens if they wish.

\item \textbf{Lend tokens:} Token lending is an additional use case implemented where a user can request a loan from the gaming platform owner. Upon approval of this loan, the owner transfers the amount desired and a record of this loan (along with interest) is recorded within the contract.

\item \textbf{Play games:} Once the user has the tokens, they can then play any of the aforementioned games and based on the result of the game, the user either pays the bank the stake they proposed or receives their winnings. The odds of winning each game are different hence the returns based on the outcome varies from game to game.

\item \textbf{Withdraw tokens:} A user can decide to withdraw their tokens from the gaming platform at any stage. When a user withdraws, their tokens get transferred back into Ether and put back onto their Ethereum account balance.
\end{enumerate}

\section{Gaming DApp Design} \label{sec:gaming-dapp-design}

We design a Decentralized Application (DApp) for playing probability games. As discussed before, the game options include Roll the Dice, Slot Machine, Roulette and Blackjack. We also include the lending use case, which allows for a novel functionality. As discussed in Section \ref{sec:related-work}, to the best of our knowledge, this functionality is currently not offered for fungible tokens by existing platforms.

This section serves to describe our design choices for modeling the use cases. To provide a deeper understanding of our ideas, we present several diagrams to explain processes and entities within our DApp. Looking at the actors involved in probability games, we first model users/players that can earn or lose tokens by placing bets on an outcome of a game. We add another type of user, a representative of the gaming platform, who can interact with the smart contract and increases the complexity of our application.
A justification of having the gaming platform as its own actor is that it is needed for the lending use case, where it actively has to approve a loan.

To establish a fair game within the contract, we make sure to collect the stakes from both parties upfront before the game starts and distribute them according to the rules of the game played. This way, we ensure that once a player and the casino agreed on the conditions and started playing the game, both actors will definitely receive their share once they win.
Additionally, we assume that there is general law enforcement outside of the contract that guarantees that users will pay back their debt since our contract does not inherently provide this functionality for lending tokens.

We choose to implement a ``gaming platform wallet'', an account with which users can manage their spending and losses within the gaming platform.
This wallet feature was chosen on the basis that there is a commonly used withdrawal pattern in Solidity programming, a practice that ensures to never let a third user initiate payments to other users. For this reason, instead of distributing token to players' accounts, only they can initiate a payment to themselves by withdrawing tokens from their gaming platform balance.
Gaming platforms generally work with their own currency, as for example gaming chips, this therefore provides another reason for tokenizing Ether and having a gaming platform account. In
addition, it simplifies lending to work with a local currency, since it is much safer for the gaming platform to lend out chips rather than actual tokens that might be spent outside of their institution.

Regarding the implementation of our application within smart contracts, we initially considered creating a modular environment in which requesting a loan and playing a game would both be part of their own contracts, as both situations work under different conditions. Additionally, we planned to have a different contract for each game type. However, for the current prototype, we decided to focus on the functionalities of our application, and to implement one big contract instead. For future projects we would choose to not work with a monolithic approach as comprehensibility and maintainability can suffer. A representation of the smart contract and its variables and functions can be seen in the class diagram in Fig. \ref{fig:casino} (here we model smart contracts as classes).

\begin{figure}[!htb]
\centering
\includegraphics[width=5.5cm]{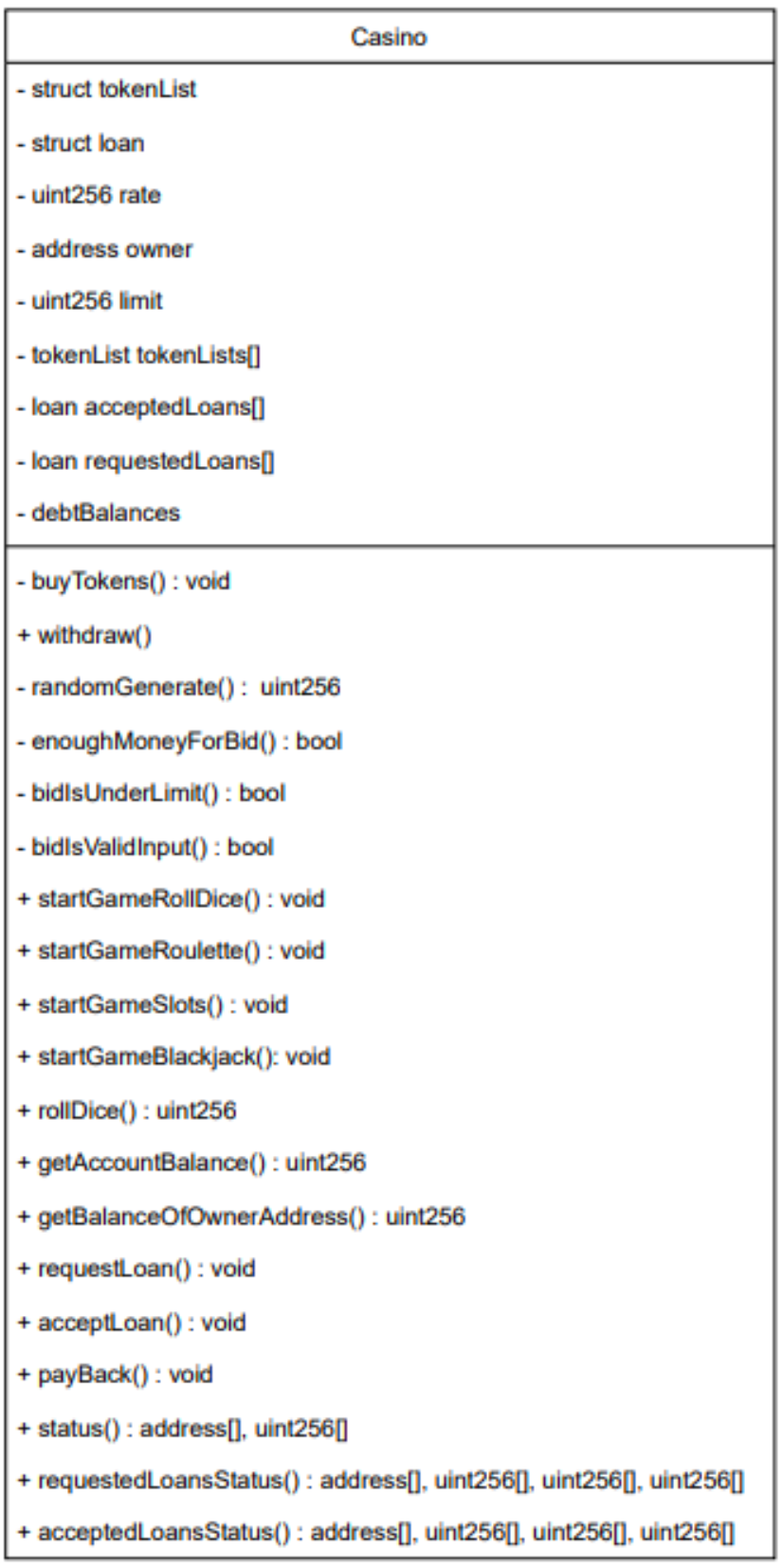}
\caption{Class Diagram: Gaming Platform}
\label{fig:casino}
\end{figure}

We also modeled a hypothetical class diagram for the Blackjack game to demonstrate how
such a smart contract could be implemented. The diagram is shown in Fig. \ref{fig:blackjack}.

\begin{figure}[!htb]
\centering
\includegraphics[width=5.5cm]{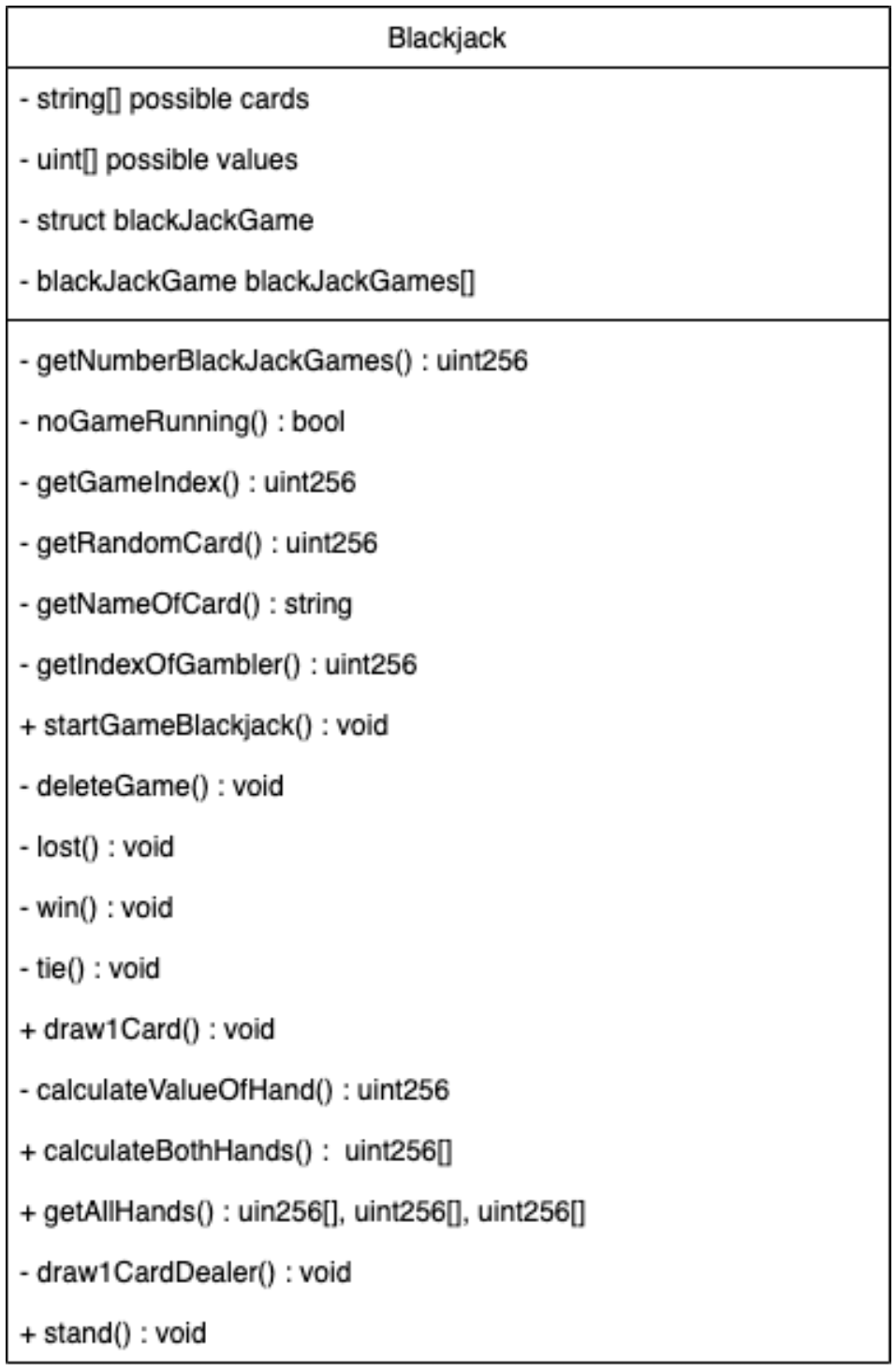}
\caption{Class Diagram: Blackjack}
\label{fig:blackjack}
\end{figure}

To develop a deeper understanding of our design choices, we also include two sequence diagrams illustrating the roll dice game (Fig. \ref{fig:rolldice}) and the lending use case (Fig. \ref{fig:lendtoken}).

\begin{figure}[!htb]
\centering
\includegraphics[width=\columnwidth]{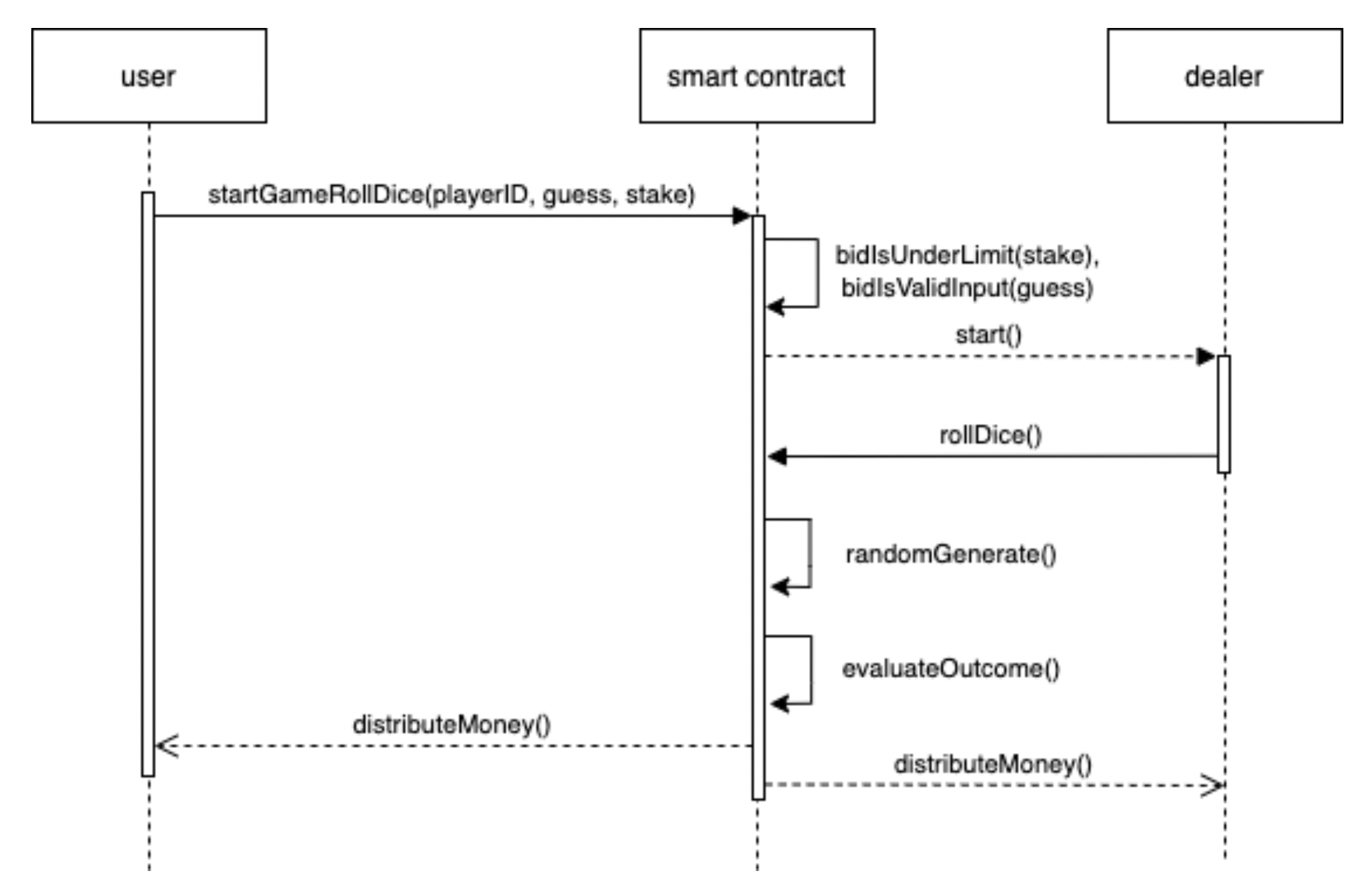}
\caption{Sequence Diagram: Roll the Dice}
\label{fig:rolldice}
\end{figure}

\begin{figure}[!htb]
\centering
\includegraphics[width=\columnwidth]{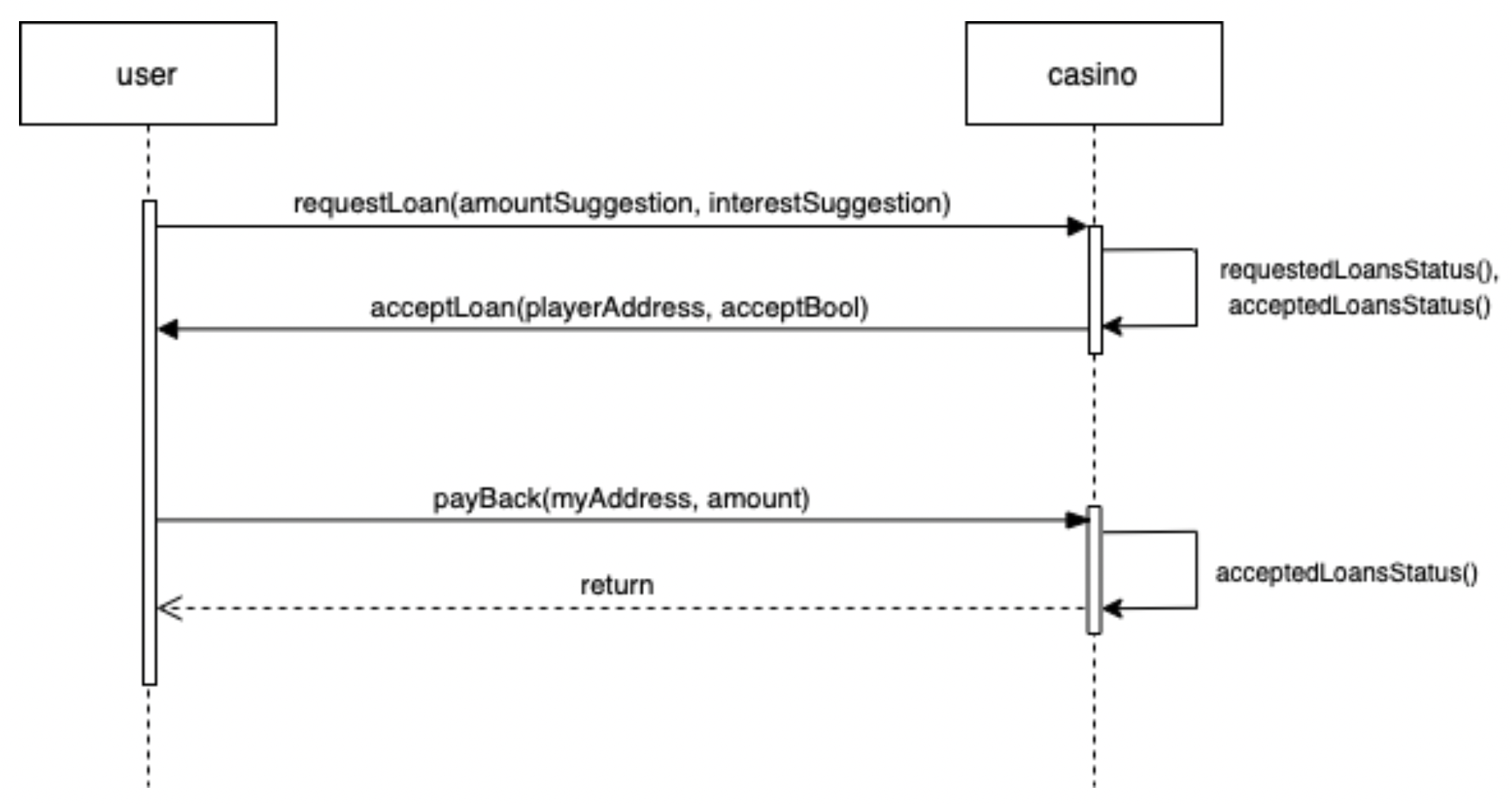}
\caption{Sequence Diagram: Lend Tokens}
\label{fig:lendtoken}
\end{figure}

Finally, we conclude this section with a state diagram that covers all the states that a user can find themselves in in terms of their financial stability (with respect to tokens).

\begin{figure}[!htb]
\centering
\includegraphics[width=\columnwidth]{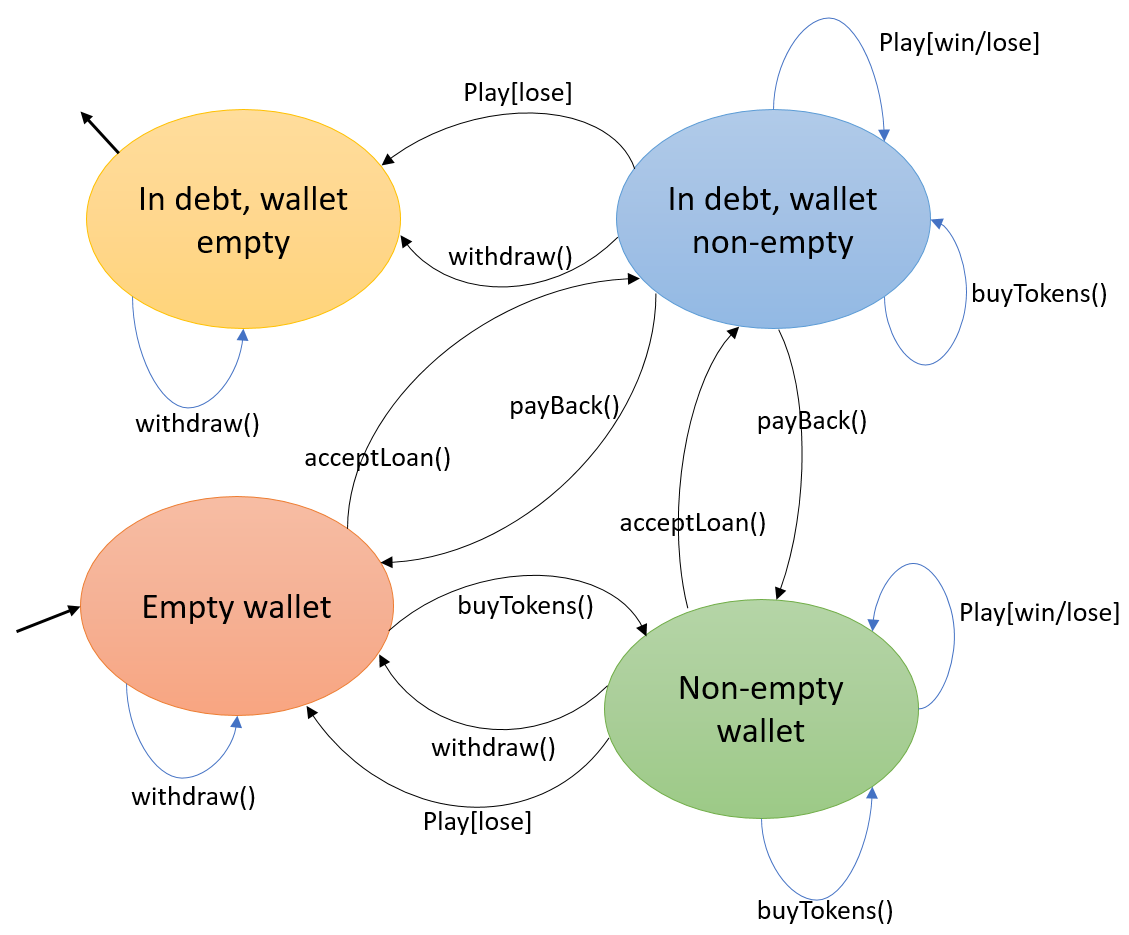}
\caption{State Diagram}
\label{fig:statediagram}
\end{figure}

\section{Implementation} \label{sec:implementation}

The front end aspect of our project provides a graphical user interface for displaying the functionalities of our Ethereum smart contract. For front-end interaction, we use React \cite{React2022, Bhalla2020} to provide a visual interface for the smart contract. React is a free and open-source front-end JavaScript library for building user interfaces using UI components. We ran our client-side application (React) in conjunction with Solidity \cite{Solidity2022} using Hardhat \cite{hardhat2022}. Similar to tools such as Ganache \cite{Ganache2022} and Truffle \cite{truffle2022}, Hardhat is used to develop an Ethereum environment and framework for full stack development. To enable interactions with our smart contract on the blockchain, we use Ether.js \cite{Ethers2022}. Similar to Web3.js \cite{web3js}, Ether.js is a javascript web client library, which we use to build our javascript frontend and interact with the Ethereum blockchain. MetaMask \cite{Meta2022} is a tool, which we used that allows users to store and manage account keys, broadcast transactions, send and receive Ether, and securely connect to decentralized applications through a web browser.

In this section, we list the Solidity source code of some of the significant functions of the gaming platform smart contract. The complete source code of the prototype is available on GitHub: https://github.com/Newman251/CasinoSmartContract.

\subsection{Buy Tokens}

This function allows anyone to buy tokens by putting Ether into the contract. At least 10 tokens have to be bought, and a maximum of 1000 tokens can be bought. If the user already has an account on the gaming platform, the newly bought tokens get added to his account. Otherwise, a new account is created for the user to which the tokens are added.

\begin{figure}[!htb]
\centering
\includegraphics[width=\columnwidth]{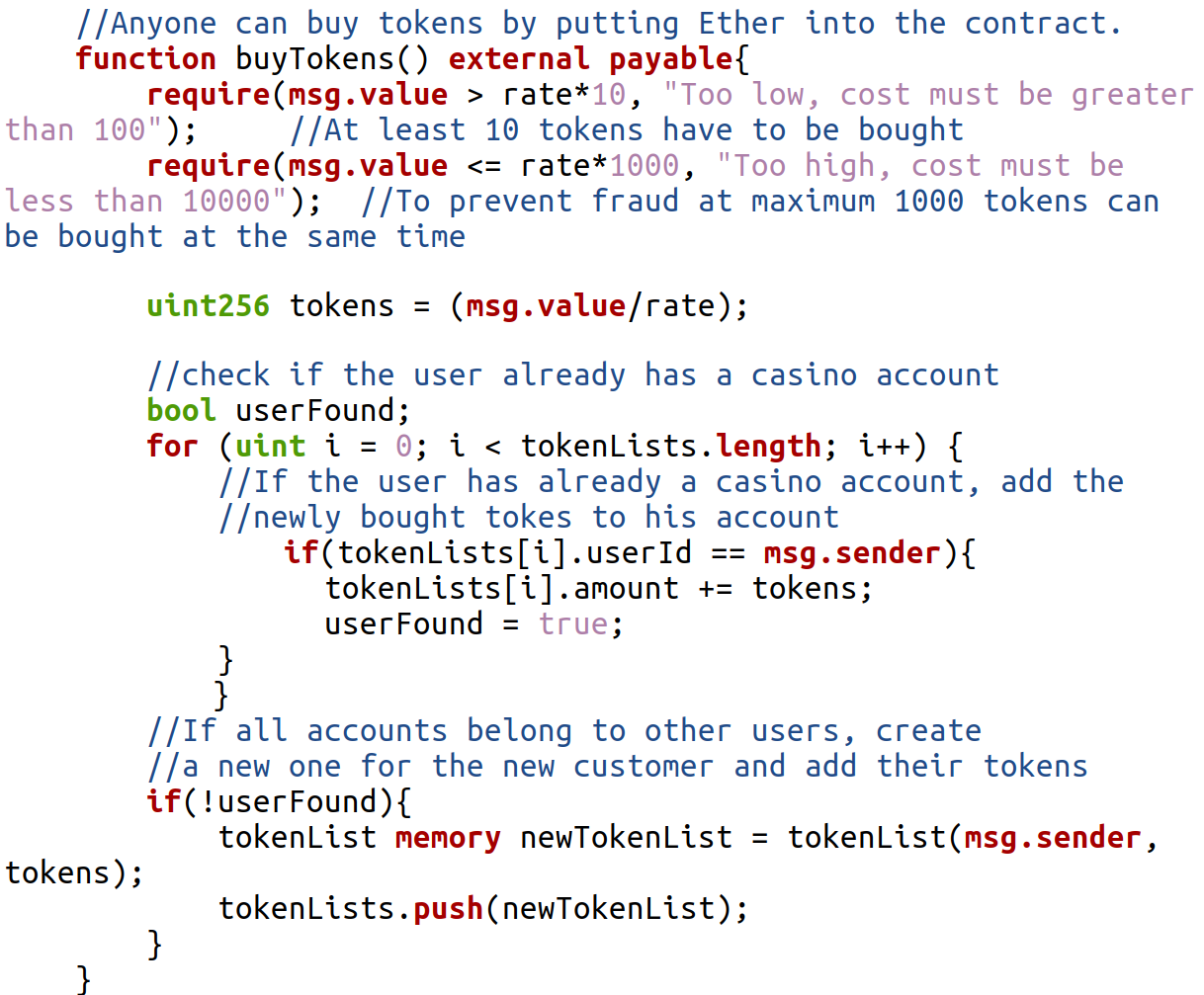}
\caption{Buy Tokens}
\label{fig:buytokens}
\end{figure}

\subsection{Request Loan}

This function allows players to request a loan by proposing an amount and interest rate. In this function, we make sure that the requested amount is not less than 100 and not more than 10000. Moreover, the interval of the proposed interest rate is also verified. The loan proposal requires approval by the gaming platform owner.

\begin{figure}[!htb]
\centering
\includegraphics[width=\columnwidth]{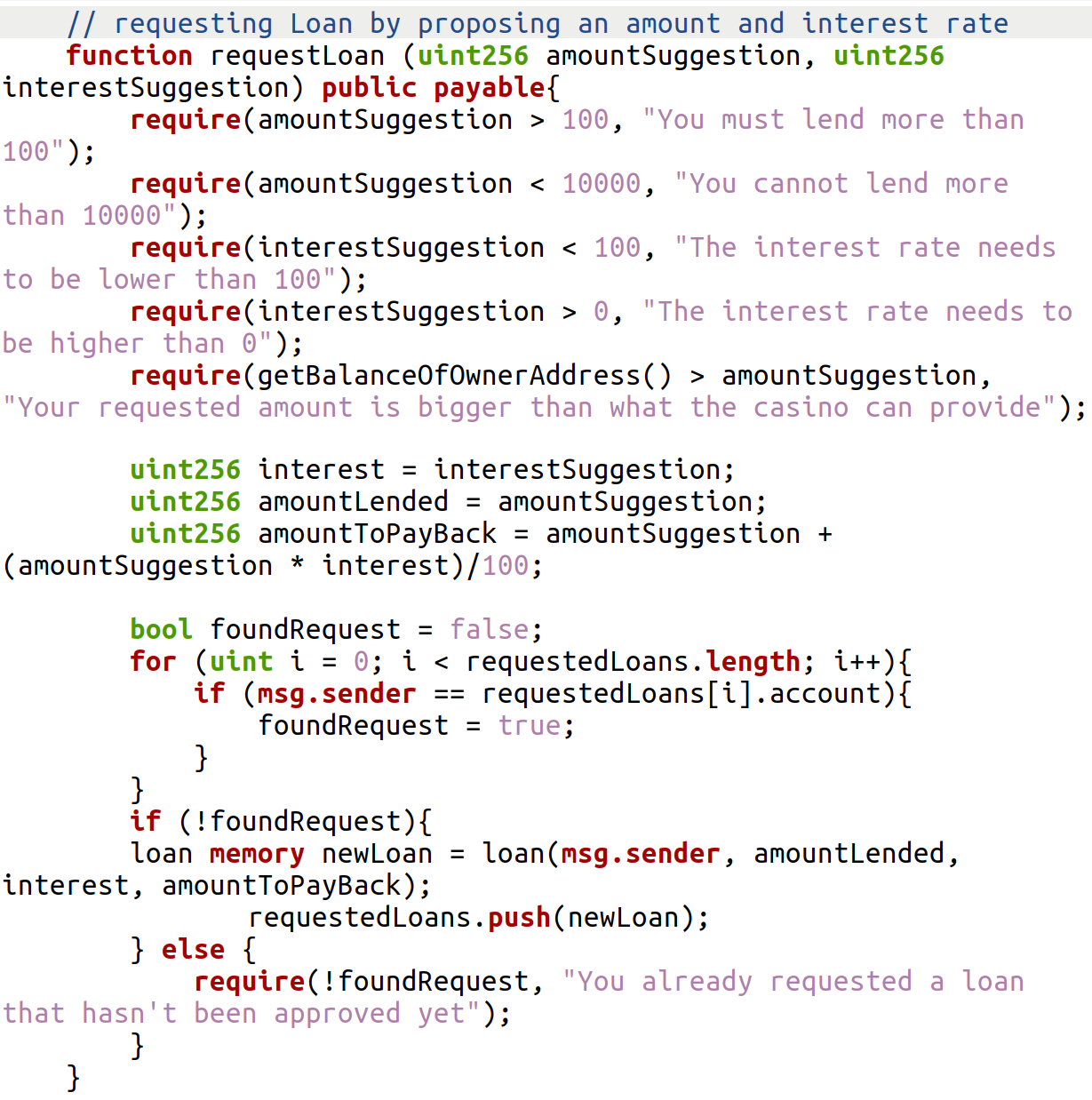}
\caption{Request Loan}
\label{fig:requestloan}
\end{figure}

\subsection{Accept Loan}

The gaming platform owner can accept a loan request by inserting the player's address and a boolean (true = accept) and (false = reject). If the loan is accepted, the smart contract transfers the balance amount to the message sender.

\begin{figure}[!htb]
\centering
\includegraphics[width=\columnwidth]{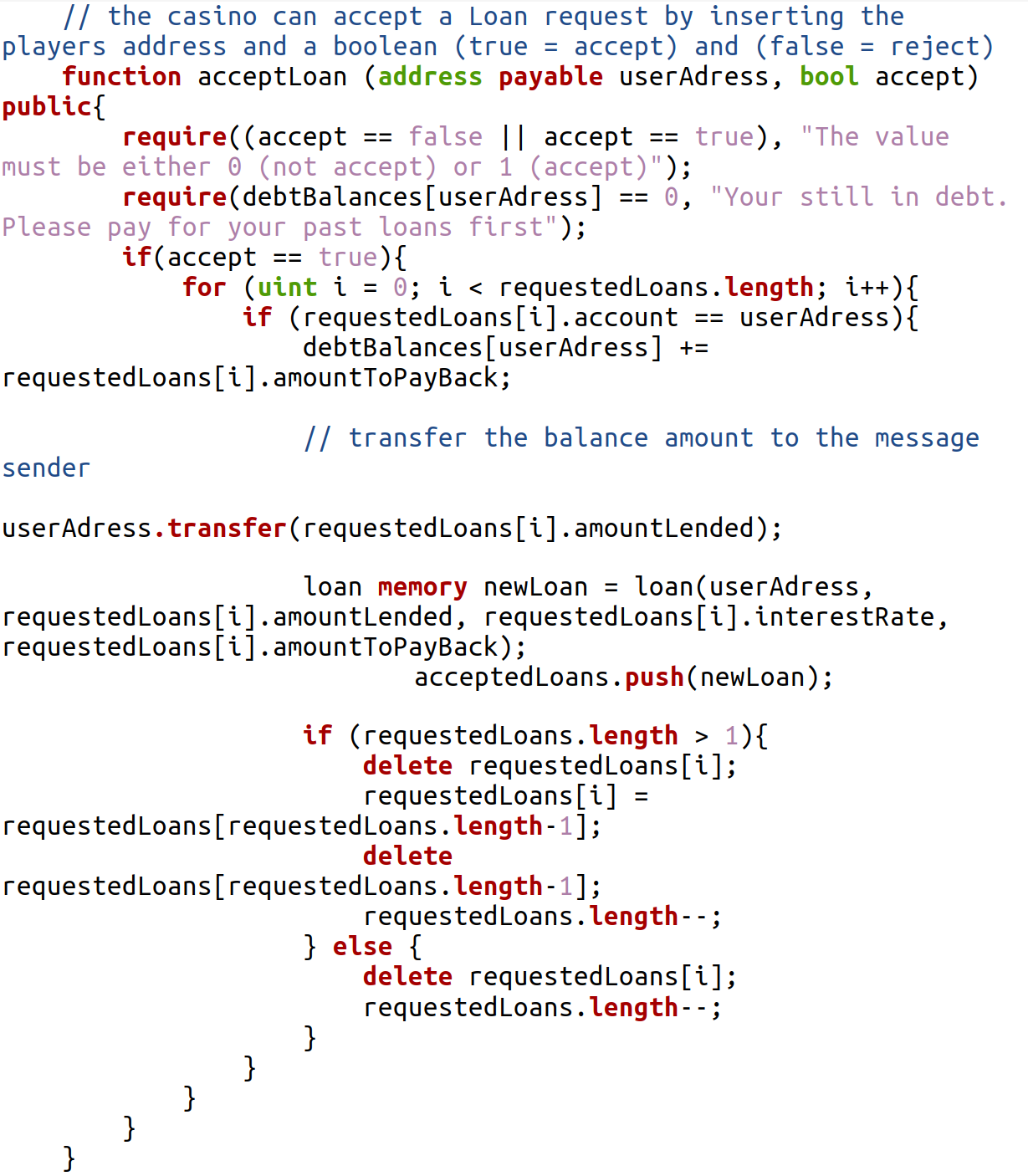}
\caption{Accept Loan}
\label{fig:acceptloan}
\end{figure}

\section{Conclusion}\label{sec:conclusion}

In this paper, we presented a decentralized online gaming platform. The smart contract of the gaming platform provides security for the players as well as the gaming platform itself. This means that all stakeholders are ensured of fairness of the rules of the games, as well as fairness regarding their account balances and winnings. Furthermore, we propose lending functionalities in this gaming platform. This allows players to engage in games using borrowed tokens in addition to playing with their own tokens. We included the implementation details of a prototype of the platform as a Decentralized Application (DApp) on the Ethereum blockchain. The complete source code of the prototype is available on GitHub.

\section*{Acknowledgment}

Wisnu Uriawan wishes to acknowledge the MORA Scholarship from the Indonesian government and the LIRIS laboratory, UMR 5205 CNRS, INSA de Lyon, which partially supports and funds his PhD work.

\bibliographystyle{./IEEEtran}
\bibliography{./IEEEabrv,./IEEEgames}

\end{document}